# Precision measurement of the Casimir-Lifshitz force in a fluid

Jeremy N. Munday
*Department of Physics, Harvard University, Cambridge, MA 02138*
Federico Capasso
*School of Engineering and Applied Sciences, Harvard University, Cambridge, MA 02138*

The Casimir force, which results from the confinement of the quantum mechanical zero-point fluctuations of the electromagnetic fields, has received significant attention in recent years for its effect on micro- and nano-scale mechanical systems. With few exceptions, experimental observations have been limited to conductive bodies interacting separated by vacuum or air. However, interesting phenomena including repulsive forces are expected to exist in certain circumstances between metals and dielectrics when the intervening medium is not vacuum. In order to better understand the effect of the Casimir force in such situations and to test the robustness of the generalized Casimir-Lifshitz theory, we have performed the first precision measurements of the Casimir force between two metals immersed in a fluid. For this situation, the measured force is attractive and is approximately 80% smaller than the force predicted by Casimir for ideal metals in vacuum. We present experimental results and find them to be consistent with Lifshitz's theory.

According to quantum electrodynamics, fluctuations of the electromagnetic fields occur in vacuum and result in a zero-point energy given by $E = \frac{1}{2}\sum \hbar \omega_i$, where the sum is evaluated over the angular frequencies $\omega_i$ of the normal modes. The introduction of two grounded, charge neutral metallic plates modifies the boundary conditions of these fluctuations, causing the fields to go to zero at the plates' surfaces. This reduces the sum of the angular frequencies by excluding some of the normal modes and alters the zero-point energy. The energy associated with this configuration as a function of plate separation can be obtained by subtracting the zero-point energy when the plates are at a finite separation $d$ from that corresponding to an infinite separation. The derivative of this energy yields a force, which was first predicted by H. G. B. Casimir in 1948 [1].

During the 1950s-60s, Lifshiftz, Dzyaloshinskii, and Pitaevskii generalized Casimir's result to included dielectrics [2, 3]. This formalism, based on the fluctuation-dissipation theorem, relies on the knowledge of the dielectric functions of the interacting materials to compute the force. In the case of small surface separations, this formalism provides a complete description of the non-retarded van der Waals interaction [4-6]. For larger separations between uncharged ideal metals, Casimir's result is obtained.

Experimental observations of the Casimir force began in the late 1950s [7, 8], although precision methods were not developed for nearly four decades [9-12]. A multitude of experimental techniques have been used to study Lifshitz's theory in the van der Waals regime (typically below 10nm surface separation) [13-17]. Measurements between metallic surfaces in water have been performed using AFM [18]; however, as with most van der Waals force measurements, little attention was given to surface roughness corrections or absolute distance determination. Other techniques have been developed for precision, long-range Casimir force measurements between metals in vacuum using MicroElectroMechanical Systems (MEMS) [12], AFM [10], and variety of torsional- and spring-based techniques for plate-plate and plate-cylinder geometries [11, 19].

More recently, attempts have been made to study the Casimir force under modified boundary conditions. Experiments have been performed between dissimilar metals [20], a metal and a semiconductor [21], a bulk metal and a thin film [22], and materials whose reflectivity in the visible range can be switched from reflective to transparent [23]. However, all of these measurements have been performed on conductive materials separated by vacuum or air.

Here we present detailed measurements of the Casimir force between metals separated by a medium other than vacuum/air. This method allows one to tune the Casimir-Lifshitz force in ways not possible in vacuum. We show that the introduction of ethanol screens the Casimir force to approximately 20% the value predicted by Casimir for ideal metals across vacuum. We further posit that it should be possible to measure repulsive Casimir-Lifshitz interactions between a metal and a dielectric immersed in fluid using this method.

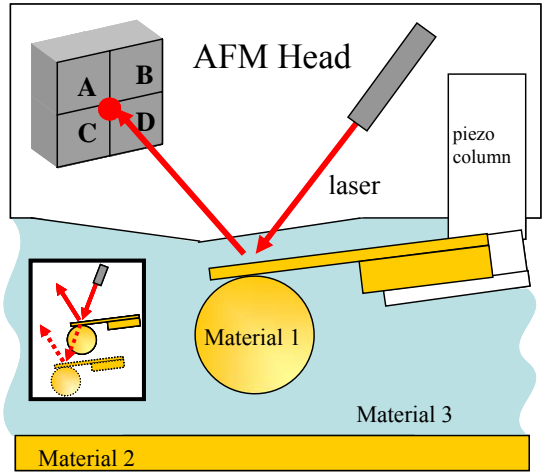

FIG. 1: (Color online) Schematic representation of the experimental setup (not to scale). A polystyrene sphere is attached to a cantilever and coated with gold. An AFM with fluid cell is used to measure surface forces as the sphere approaches the bottom plate. Inset shows a schematic of the relative motion of the cantilever with respect to the laser, which is determined and removed.

A schematic of the experimental setup is shown in Figure 1. A $42.7 \pm 0.2 \mu m$ diameter polystyrene sphere is attached to a cantilever used for AFM imaging (NovaScan). A 100nm gold layer is then evaporated onto both sides of the cantilever to create a metallic front surface for the Casimir force interaction and a reflective backside necessary for the laser deflection measurement technique. Nearly equal film thicknesses are necessary on both sides of the cantilever in order to reduce cantilever curvature due to material stresses induced at the surface interface. An additional 10nm of gold is evaporated at sharp angles with respect to the cantilever in order to ensure electrical continuity of the evaporated film and the metallized sphere. The cantilever is inserted into a commercially available AFM (Asylum Research MFP-3D) containing a fluid cell for measurements.

Prior to setup, standard cleaning procedures are performed on all surfaces. The gold plate and fluid cell are ultrasonically cleaned for 30 minutes in ethanol followed by drying in nitrogen airflow. The cantilever chip is similarly rinsed with ethanol and nitrogen without ultrasonic cleaning to avoid damaging the cantilever. The liquid ethanol used for the experiment contains <0.1% $H_2O$ (Sigma-Aldrich) and is filtered through a 0.2μm PTFE filter. Once assembled, the experiment is allowed to equilibrate for one hour prior to measurements. For all experiments, the cantilever is completely submerged and no evidence of micro-bubbles or particle contamination is seen.

A laser (1mW at λ=860nm) is reflected off the cantilever into a four-quadrant photodetector for monitoring the cantilever deflection (Fig. 1). Any vertical displacement or bending of the cantilever can be detected through the difference signal between the top two quadrants and the bottom two quadrants of the photodetector. A piezoelectric column drives the cantilever toward the plate, which is monitored using a Linear Variable Differential Transformer (Asylum Research) to ensure accurate detection of displacements and to avoid nonlinearities and hysteresis inherent to piezoelectrics. The interaction between the sphere and the plate is then detected through the motion of the cantilever as the sphere approaches the plate. The inset of Fig. 1 depicts the relative motion of the cantilever with respect to the laser, which results in an artificial deflection signal. This signal was determined to be independent of surface separation and is removed from the final data.

The photodetector voltage signal is converted into a force signal through calibration with a known force. The deflection of the cantilever obeys Hooke's law: $F_{spring} = -k\, d_{cantilever}$, where $d_{cantilever}$ is the distance the tip of the cantilever has bent and $k$ is the spring constant of the cantilever. When an external force is applied to the cantilever, an equilibrium condition is reached when $F_{external} = k\, d_{cantilever}$. In order to determine the deflection distance $d_{cantilever}$ from the photodetector signal, the piezoelectric column moves the cantilever toward the surface of the plate until the sphere and plate are in contact, and the normal force of the plate against the sphere causes a linear deflection of the cantilever versus piezo displacement. The slope of this linear contact region, $m$, can be used to convert the photodetector signal $V_{det}$ into the distance the cantilever has bent, $d_{cantilever} = -\frac{V_{det}}{m}$. The force on the cantilever is $F_{external} = k \; d_{cantilever} = -k \frac{V_{det}}{m} = \alpha V_{det}$, where $\alpha$ is the force constant, which converts the raw photodetector voltage signal into a force signal. To determine $\alpha$, a known force is applied between the plate and the sphere, and $\alpha$ is determined from a fit to this force.

For high precision Casimir force measurements, an electrostatic force is typically used to obtain the spring constant [10]; however, when the intervening material is not vacuum, dielectric screening can reduce the electrostatic force by over an order of magnitude. For ethanol the reduction factor is $\varepsilon_{ethanol} = 24.3$, the dielectric constant. Thus, a different method is needed to determine the spring constant. Other common methods for determining the spring constant without making a direct force measurement (e.g. thermal [24], added mass [25] or Sader [26] methods) are considerably less accurate. To avoid this problem, we have adapted a method which uses a hydrodynamic force rather than an electrostatic force to determine the force constant [27]. The hydrodynamic force between a sphere and a plate separated by a distance $d$ is given, in the limit of $R >> d$, by [28, 29]:

$$F_{hydrodynamic} = -\frac{6\pi\eta v}{d}R^2, \qquad (1)$$

where $R$ is the radius, $\eta$ is the fluid viscosity and $v$ is the velocity of the sphere (in our convention this is negative when moving toward the plate). For a given sphere and fluid, the piezo velocity can be varied to provide a large hydrodynamic force for force constant determination or a negligibly small force for observation of the Casimir-Lifshitz force.

We typically take data at several piezo velocities for analysis, using a feedback loop to ensure a constant tip velocity. Figure 2 shows the photodetector signal versus piezo displacement for three different velocities collected at a sampling rate of 5kHz. For all three velocities, the hydrodynamic force is the dominant interaction at surface separations larger than 200nm. At a speed of 0.9μm/s, the Casimir-Lifshitz force becomes comparable to the hydrodynamic force at a separation of approximately 50nm and causes the photodetector signal to change from positive to negative, corresponding to an attractive force at small separations.

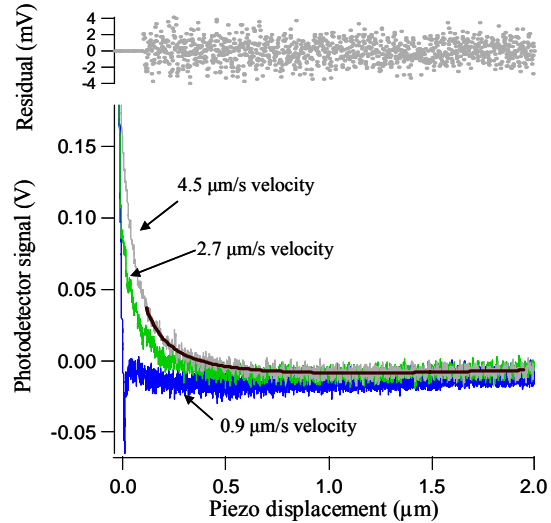

FIG. 2: (Color online) Raw deflection data versus piezo displacement (thin lines) at various approach speeds, and corresponding fit (thick black line) to the hydrodynamic force (4.5 μm/s tip velocity) used for determination of the force constant and absolute surface separation. Residual shows no systematic error in the least-squares curve fit.

The force constant $\alpha$ and the actual sphere-plate separation at contact $d_0$ are determined by fitting the cantilever deflection data for large sphere velocity to Eq[1] with $d = d_{piezo} + d_{cantilever} + d_0$, where $d_{piezo}$ is the piezo displacement. This procedure is performed on the raw data with a piezo velocity of 4.5μm/s (Fig. 2) for distances greater than 200nm, where the hydrodynamic force is the dominant interaction, and there is no significant modification to the force due to surface roughness. The residual in the plot shows no systematic error in the least-squares curve fit. The only two free variables, $\alpha$ and $d_0$, are determined to be $5.32 \pm 0.07$ nN/V and $20 \pm 3$ nm, respectively.

The Casimir-Lifshitz force between the two gold surfaces separated by ethanol is then measured using a slow piezo approach speed. We collected data for 11 consecutive runs at an approach speed of 45nm/s. A sampling rate of 5kHz is used to acquire over 100 points per nanometer, which are then averaged to give one data point per nanometer. Although the piezo distance is known to <1nm, this averaging range is chosen due to the scatter in the data on this distance scale. No noticeable difference is found using a slightly larger or smaller range.

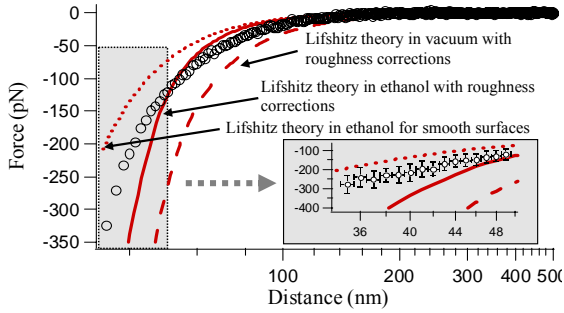

FIG. 3: (Color online) Experimental results (circles) with three different calculations based on theory: Lifshitz's theory for two perfectly smooth gold surfaces separated by ethanol (dotted line), Lifshitz's theory for gold surfaces in ethanol with surface roughness corrections (solid line) and Lifshitz's theory for gold surfaces in vacuum with roughness corrections (dashed line). Inset: data from 35-50nm showing the deviation from theory below 50nm. The data points and error bars

In addition to the Casimir-Lifshitz force, residual electrostatic and hydrodynamic forces are still present but are small by comparison. The contact potential between the plate and the sphere in air is $V_0 = 130\text{mV}$ as determined by varying the bias voltage on the sphere while keeping the plate grounded. The resulting electrostatic force between the sphere and plate in ethanol at a separation of 40 nm is -10pN. With a velocity of 45nm/s, the hydrodynamic drag force at this separation is 12pN. The total residual force is given by

$$F_{electrostatic} + F_{hydrodynamic} = -\left(\frac{\varepsilon_0 V_0^2}{\varepsilon_{ethanol}} + 6\eta v R\right)\frac{\pi R}{d}$$

and is 2pN at 40nm compared to the Casimir-Lifshitz force of -260pN.

The experimental data is compared to Lifshitz theory for a gold plate interacting with a gold sphere immersed in liquid ethanol. For three materials, the interaction force can be written as [2]:

$$F_{C-L}(d) = \frac{\hbar}{2\pi c^2} R \int_{\xi=0}^{\infty} \int_{p=1}^{\infty} \varepsilon_3 \, p \, \xi^2 \{\ln[1 - \Delta_{31}^{(1)} \Delta_{32}^{(1)} e^{-x}]$$
$$+ \ln[1 - \Delta_{31}^{(2)} \Delta_{32}^{(2)} e^{-x}]\} dp \, d\xi \,, \quad (2)$$

where

$$\Delta_{3k}^{(1)} = \frac{s_k \varepsilon_3 - s_3 \varepsilon_k}{s_k \varepsilon_3 + s_3 \varepsilon_k}, \quad \Delta_{3k}^{(2)} = \frac{s_k - s_3}{s_k + s_3},$$

$$x = \frac{2d\sqrt{\varepsilon_3}\,\xi\, p}{c}, \quad s_k = \sqrt{p^2 - 1 + \frac{\varepsilon_k}{\varepsilon_3}},$$

$\hbar$ and $c$ are the usual fundamental constants, and $\varepsilon_1$, $\varepsilon_2$, and $\varepsilon_3$ are the dielectric functions of the sphere, the plate, and the intervening medium, respectively, evaluated at imaginary frequencies $i\xi$ according to:

$$\varepsilon_k \equiv \varepsilon_k(i\xi) = 1 + \frac{2}{\pi}\int_{x=0}^{\infty} \frac{x\,\text{Im}[\varepsilon_k(x)]}{x^2 + \xi^2}\,dx. \quad (3)$$

The dielectric functions are obtained from [30, 31]. It was recently shown that the calculated force between two metals in vacuum can vary by as much as 5% due to the variation in the optical properties of gold which occur for different samples [32]. The inclusion of a third material (ethanol), whose optical properties are less well known, should further increase the theoretical uncertainty to above 5%. Surface roughness further modifies the calculation of the Casimir-Lifshitz force. The total force including this correction can be written as [33]:

$$F(d) = \sum_{i,j} v_i^{(sp)} v_j^{(pl)} F_{C-L}\left(d - \left(\delta_i^{(sp)} + \delta_j^{(pl)}\right)\right), \quad (4)$$

where $v_i$ is the fraction of the surface area of the sphere (sp) or plate (pl) displaced a distance $\delta_i$ from an ideally smooth surface. The values for $v_i$ as a function of $\delta_i$ are measured using an optical profiler over a 2μm x 2μm section [34].

Figure 3 shows the comparison between our experimental data and theory. The calculation of the Casimir force for ideal metals (not shown) overestimates it by a factor of ~4.5, while Lifshitz's theory for gold surfaces in ethanol without the roughness correction (dotted line) generally underestimates the measured force.

Inclusion of roughness effects (solid line) gives a better fit to the data; however, the discrepancy increases at separations below 50nm (inset). The magnitude of the measured force appears to be approximately 5-10pN larger than the theory in the range 50-100nm and is consistently smaller below 50nm, where the roughness corrections begin to breakdown. For comparison, Lifshitz's theory including roughness corrections for two gold surfaces separated by vacuum (dashed line) is plot in Fig. 3 and is larger than the measured force by a factor of ~2.

While the force constant $\alpha$ has an error of <2% and the theory is believed to be accurate to within ~5% as previously discussed, comparison between theory and experiment cannot be made at this level. Because of the approximate $1/d^3$ dependence of the Casimir-Lifshitz force, the comparison is limited by the uncertainty in $d_0$ [35]. For our experiment, $\delta d_0 = \pm 3$ nm, which corresponds to a 25% uncertainty in the comparison at a separation of 50nm. To achieve a level of 2%, the uncertainty in $d_0$ would need to be reduced to a few angstroms. This is the leading source of error in the comparison at small separations and is often overlooked in the literature (for a critic of related literature see Iannuzzi *et al.* [35]).

Calculations have recently shown that repulsive interactions may be achieved using a metallized sphere above a silica plate immersed in ethanol [36]. Repulsive forces could be of great importance for technological applications by counter-balancing gravity to suspend one surface above another at close range using no classical electromagnetic forces. For a 100μm radius gold sphere 50nm above a silica plate, the resulting repulsive force is 135pN. Using a technique similar to the one presented here a measurement of this long-range quantum electrodynamical repulsion should be measurable.

In conclusion, we have conducted the first precision measurements of the Casimir-Lifshitz force between two metal surfaces (gold) separated by a fluid (ethanol). The results were found to be consistent with Lifshitz's theory, and errors were discussed. A straightforward extension of this methodology could be used to study the Casimir-Lifshitz force between metals and dielectrics in fluid, which raises the intriguing possibility of achieving quantum flotation using repulsive QED forces.

The authors would like to acknowledge M. B. Romanowsky, N. Geisse, K. Parker, and D. Iannuzzi for helpful discussions. This project was partially supported by NSEC, under NSF contract number PHY-0117795 and by the Center for Nanoscale Systems at Harvard University. JNM gratefully acknowledges financial support from the NSF Graduate Research Fellowship Program (GRFP).